\documentclass[12pt,a4paper,final]{iopart}

\usepackage{iopams}
\usepackage{graphicx}
\usepackage{dcolumn}
\usepackage{float}

\begin{document}

\title{\boldmath  Analysis of optical data using extended Drude model and generalized Allen's formulas \unboldmath}

\author{Jungseek Hwang}
\address{Department of Physics, Sungkyunkwan University, Suwon, Gyeonggi-do 16419, Republic of Korea\\ Photon Sciences, Brookhaven National Laboratory, Upton, New York 11973, USA}

\ead{jungseek@skku.edu}

\date{\today}

\begin{abstract}

Extended Drude model formalism has been successfully utilized for analyzing optical spectra of strongly correlated electron systems including heavy-fermion systems and high-$T_c$ superconducting iron pnictides and cuprates. Furthermore, generalized Allen's formulas has been developed and applied to extract the electron-boson spectral density function from measured optical data of high temperature superconductors including cuprates in various material phases. Here we used a reverse process to obtain various optical quantities starting from two typical electron-boson spectral density model functions for three intriguing (normal, pseudogap, and $d$-wave superconducting) material phases in cuprates. We also assigned the calculated optical results to designated regions in the phase diagram of hole-doped cuprates and compared them with the corresponding measured optical spectra of Bi$_2$Sr$_2$CaCu$_2$O$_{8+\delta}$ (Bi-2212). This comparison suggested that this way of optical data analysis can be a convincing method to study correlated electrons in the copper oxide superconductors and other superconducting systems as well.

\end{abstract}

\maketitle

\section{Introduction}

Strongly correlated electron systems show intriguing physical properties including pseudogap and superconductivity. Electrons in these systems can be described by an extended Drude model, which has an analogous form with the dynamical magnetic susceptibility in dilute magnetic alloys\cite{mori:1965,gotze:1972}. In this model, the memory (or relaxation) function\cite{gotze:1972} was introduced and later also has been known as the optical self-energy\cite{hwang:2004}, which is closely related to the well-known quasiparticle self-energy\cite{carbotte:2005,hwang:2007b}. Therefore, the optical self-energy can carry information on the correlation like the quasiparticle self-energy. Furthermore, since the optical self-energy can carry information on both filled and empty electronic states it can be called a two-particle self-energy. The extended Drude model is also known as one-component (or one-band) model\cite{tanner:1992}. In an extreme case, the one-band can be divided into two distinct (upper Hubbard and lower Hubbard) bands by a strong Coulomb repulsion or in a strongly correlated regime\cite{kotlier:2004}. Intermediately correlated systems will show that the one-band evolves into two parts: the coherent and incoherent parts. Researchers have applied the extended Drude model formalism to various strongly correlated electron systems in normal and superconducting states and obtained a lot of important and interesting results\cite{hwang:2004,webb:1986,sulewski:1988,thomas:1988,awasthi:1993,puchkov:1996,dordevic:2001,tran:2002}

Usually one can obtain the electron-boson spectral density function from a measured reflectance spectrum through a well-developed process, which consists of a series of steps and is called as a normal process\cite{collins:1989,carbotte:1990,schachinger:2000,hwang:2006,hwang:2007}. One can get a reflectance spectrum and other optical constants starting from a given electron-boson spectral density function through an opposite order of the normal process. Through this reverse process\cite{hwang:2015a} one can learn or figure out how some characteristic features in the electron-boson density function appear in other optical constants including the reflectance spectrum\cite{hwang:2015a}. One can fit measured optical data of multi-band correlated systems using this reverse process\cite{hwang:2016}.

In this paper we obtained various optical quantities by performing the reverse process\cite{hwang:2015a} with typical model electron-boson spectral functions for three interesting (normal, pseudogap, and $d$-wave superconducting) material states and showed that characteristic features related to the model electron-boson spectral functions and the pseudogap and superconducting gap appear in the obtained optical quantities. Then we assigned the calculated results to five designated regions in the generic phase diagram of hole-doped cuprates\cite{basov:2005} by considering the observed temperature- and doping-dependent properties of the electron-boson spectral function and the related material phases\cite{hwang:2004,hwang:2006,hwang:2007,dai:1999,fong:2000,zasadzinski:2006,zasadzinski:2011,carbotte:2011}. Furthermore, we compare the assigned optical quantities with measured optical data of Bi-2212 cuprate systems in the same designated regions. This comparison shows that this approach of optical data analysis is a convincing way, which can be used to extract information on correlation between electrons in cuprates and other superconducting materials as well.

\section{Theoretical formalism}

We introduce the reverse process\cite{hwang:2015a} in some detail for the three (normal, pseudogap, and $d$-wave superconducting) material phases. It is worth to be noted that the normal phase is a non-Fermi liquid (or a strange metal) in the phase diagram of cuprates. Here we consider an interaction between electrons by exchanging mediating bosons; the interaction can be described by the electron-boson spectral density function, $I^2 B(\omega)$, where $I$ is the coupling constant between the mediating boson and an electron and $B(\omega)$ is the boson spectral density. Starting from a model $I^2 B(\omega)$ we calculate the imaginary part of the complex optical self-energy ($\tilde{\Sigma}^{op}(\omega) \equiv \Sigma_1^{op}(\omega) + i\Sigma_2^{op}(\omega)$) using the generalized Allen's formalisms\cite{allen:1971} as follows:
\begin{equation}\label{eq1}
-2\Sigma_2^{op}(\omega,T) =  \int^{\infty}_{0} d\Omega \: I^2 B(\Omega,T) \: K(\omega, \Omega, T)  + \Gamma^{op}_{imp}(\omega,T),
\end{equation}
where $K(\omega, \Omega,T)$ is the kernel of the generalized Allen's integral equation and $\Gamma^{op}_{imp}(\omega,T)$ is the optical impurity scattering rate. The kernels for the pseudogap (PG)\cite{sharapov:2005} and $d$-wave superconducting\cite{allen:1971,schachinger:2006} states can be described as follows:
\begin{equation}\label{eq2}
K(\omega, \Omega,T) \left\{
 \begin{array}{ccc}
   =& \frac{2\pi}{\omega}\int_{-\infty}^{+\infty} dz \: \frac{N(z-\Omega)+N(-z+\Omega)}{2}& \\
    & \times [n_B(\Omega)+1-n_F(z-\Omega)]& \\
    & \times [n_F(z-\omega)-n_F(z+\omega)]& \:  \:\:\:\: (\mbox{PG state}) \\
   =& \frac{2\pi}{\omega}(\omega-\Omega)\Big{\langle}\Theta(\omega-2\Delta(\theta,T)-\Omega)& \\
    & \times E\Big{(} \sqrt{1-\Big{[}\frac{2 \Delta(\theta,T)}{\omega-\Omega}\Big{]}^2} \Big{)}
  \Big{\rangle}_{\theta}& \:\:\:\: (\mbox{$d$-wave SC}),
  \end{array}
  \right.
\end{equation}
where $N(z) \equiv \tilde{N}(z)/\tilde{N}(0)$ is a normalized density of states of the quasiparticles, $\tilde{N}(z)$ is the density of states and $\tilde{N}(0)$ is the density of states at the Fermi level. $n_B(\Omega)$ $[= 1/(e^{\beta \Omega}-1)]$ and $n_F(z)$ $[= 1/(e^{\beta z}+1)]$ are the boson and fermion occupation functions, respectively and $\beta \equiv 1/(k_B T )$, $k_B$ and $T$ are the Boltzmann constant and the absolute temperature, respectively. $\Theta(z)$ is the Heaviside step function, $E(z)$ is the complete elliptic integral of the second kind, and $\langle\cdots\rangle_{\theta}$ stands for the angular average over $\theta\in[0, \pi/4]$. $\Delta(\theta, T)$ [=$\Delta(T) \cos(2\theta)$] is the $d$-wave superconducting gap, $\Delta(T)$ is the temperature-dependent superconducting energy gap with the maximum gap $\Delta_0$ at $T =$ 0. It should be noted that when $N(|z|) =$ 1 the PG phase becomes the normal state, which has a constant density of states.

From now on, we will consider $T =$ 0 case for simplicity without losing the generality. At $T =$ 0 case the PG kernel can be described as follows\cite{mitrovic:1985}:
\begin{equation}\label{eq2a}
 K(\omega,\Omega) = \left\{
 \begin{array} {ccc}
 &\frac{2 \pi}{\omega} \Theta(\omega-\Omega)(\omega-\Omega),&\:\:(\mbox{normal state}) \\
 &\frac{2 \pi}{\omega}\int^{\omega-\Omega}_0 dz \: \frac{N(z)+N(-z)}{2},& (\mbox{PG state}).
  \end{array}
  \right.
\end{equation}
Here we used the PG function, which was introduced previously\cite{hwang:2008,hwang:2008b}, as follows:
\begin{equation}\label{eq2b}
N(z) = \left\{
 \begin{array} {ccc}
 & N_0 +(1 - N_0)\Big{(} \frac{z}{\Delta_{PG}} \Big{)}^2\:\:,\:\:&(0 \leq |z| < \Delta_{PG})  \\
 & \frac{5-2 N_0}{3}\:\:, \:\:\:\:\:\:\:\:\:\:\:\:\:\:\:\:\:\:&(\Delta_{PG} \leq |z| < 2 \Delta_{PG}) \\
 & 1 \:\:, \:\:\:\:\:\:\:\:\:\:\:\:\:\:\:\:\:\:\:\:\:\:\:\:\:\:\:\:\:\:\:\:&(2\Delta_{PG} \leq |z|),
 \end{array}
 \right.
\end{equation}
where $\Delta_{PG}$ is the size of the pseudogap and $N_0$ is the normalized density of states at the Fermi level. Therefore, $1-N_0$ is the depth of the pseudogap at the Fermi level; when $N_0 = 1.0$ there is no pseudogap i.e., the system becomes the normal state, and when $N_0 = 0.0$ the pseudogap is fully developed.

The impurity scattering rate at $T =$ 0 can be described as follows:
\begin{equation}\label{eq3}
\Gamma^{op}_{imp}(\omega) = \left\{
  \begin{array} {ccc}
  & \!\!\Gamma_{imp} \:\: &(\mbox{normal state}) \\
  & \!\!\Gamma_{imp}\frac{1}{\omega}\int^{\omega}_0 dz \: \frac{N(z)+N(-z)}{2}\:   &(\mbox{PG state}) \\
  & \!\!\Gamma_{imp} \Big{\langle} E\Big{(} \sqrt{1\!-\!\!\Big{[}\frac{2 \Delta(\theta)}{\omega}\Big{]}^2} \Big{)} \Big{\rangle}_{\theta} &(\mbox{$d$-wave SC}),
  \end{array}
  \right.
\end{equation}
where $\Gamma_{imp}$ is the impurity scattering rate, which is a constant. The impurity scattering rate in the SC state strongly depends on frequency near twice of the $d$-wave superconducting gap, 2$\Delta(\theta)$ [=2$\Delta_0\cos(2\theta)$]. In the PG phase, the impurity scattering rate at the Fermi level, $\Gamma^{op}_{imp}(0) = N_0 \Gamma_{imp}$. It should be noted that if $T \neq$ 0 the temperature-dependence will appear in $N(z,T)$ for the PG case and $\Delta(\theta,T)$ for the SC state.

After obtaining the imaginary part of the optical self-energy using Eq. (1) in a wide enough spectral range we calculate the real part of the optical self-energy with the obtained imaginary part using the Kramers-Kronig relation between them as follows:
\begin{equation}\label{eq4}
-2\Sigma_1^{op}(\omega) = -\frac{2\omega}{\pi}P\int^{\infty}_0d\omega'  \: \frac{[-2\Sigma_2^{op}(\omega')]}{\omega'^2-\omega^2},
\end{equation}
where $P$ stands for the principal part of the improper integral. We also calculate the mass enhancement function [$m_{op}^*(\omega)/m_b \equiv \lambda^{op}(\omega)+1$] using the relation between $m_{op}^*(\omega)$ and $-2\Sigma_1^{op}(\omega)$ i.e., $m_{op}^*(\omega)/m_b = -2\Sigma_1^{op}(\omega)/\omega + 1$, where $m_{op}^*(\omega)$ is the optical effective mass and $m_b$ is the band mass.

Then we calculate the complex optical conductivity ($\tilde{\sigma}(\omega) \equiv \sigma_1(\omega)+i\sigma_2(\omega)$) in the extended Drude model formalism\cite{gotze:1972,hwang:2004} as follows:
\begin{equation}\label{eq5}
\tilde{\sigma}(\omega)=\frac{i}{4\pi}\frac{\omega_p^2}{\omega+[-2\tilde{\Sigma}^{op}(\omega)]},
\end{equation}
where $\omega_p$ is the plasma frequency. The plasma frequency is the first parameter of the two parameters, which we need to provide in the reverse process. The plasma frequency square is proportional to the charge carrier density in the single band of interest. We can get any other optical constants including reflectance spectrum using the relationships between them\cite{wooten}. The complex optical dielectric function ($\tilde{\epsilon}(\omega) \equiv \epsilon_1(\omega)+i\epsilon_2(\omega)$) can be obtained from the complex optical conductivity using the following equation,
\begin{equation}\label{eq6}
\tilde{\epsilon}(\omega) = \epsilon_H + i\frac{4 \pi}{\omega}\tilde{\sigma}(\omega),
\end{equation}
where $\epsilon_H$ is the background dielectric constant, which may come from contributions of any bands other than the single band located at high frequency region. The background dielectric constant is the second parameter of the two parameters, which we need to provide in the reverse process. Another interesting quantity, the energy loss function, can be described in terms of the dielectric function as follows:
\begin{equation}\label{eq6a}
Loss(\omega) \equiv Im\Big{[}\frac{-1}{\tilde{\epsilon}(\omega)}\Big{]} = \frac{\epsilon_2(\omega)}{[\epsilon_1(\omega)]^2+[\epsilon_2(\omega)]^2}.
\end{equation}
The normal incidence reflectance ($R(\omega)$) can be obtained using a well-known Fresnel's equation in terms of the dielectric function as follows:
\begin{equation}\label{eq7}
R(\omega) = \Big{|}\frac{\sqrt{\tilde{\epsilon}(\omega)}-1}{\sqrt{\tilde{\epsilon}(\omega)}+1}\Big{|}^2.
\end{equation}

For SC state the complex optical conductivity ($\tilde{\sigma}^{SC}(\omega) = \sigma_1^{SC}(\omega) +i \sigma_2^{SC}(\omega)$) can be decomposed into two components as follows:
\begin{eqnarray}\label{eq8}
\sigma_1^{SC}(\omega)&=& \frac{\omega_{sp}^2}{8}\delta(\omega) + \sigma_1^{res}(\omega) \nonumber \\
\sigma_2^{SC}(\omega)&=& \frac{\omega_{sp}^2}{4\pi \:\omega} + \sigma_2^{res}(\omega),
\end{eqnarray}
where $\omega_{sp}$ is the superfluid plasma frequency, which, in principle, should be smaller than the plasma frequency, $\omega_p$. $\delta(\omega)$ is the Dirac delta function. $\sigma_1^{res}(\omega)$ and $\sigma_2^{res}(\omega)$ are the residual optical conductivities, which do not get involved in the superfluidity. It is worth to be noted that in a finite frequency region $\sigma_1^{SC}(\omega)$ is identical to $\sigma_1^{res}(\omega)$. Therefore, using optical spectroscopy techniques one can access only $\sigma_1^{res}(\omega)$ part of the full $\sigma_1^{SC}(\omega)$ since one can take optical spectra only in a finite frequency region and the $\sigma_2^{res}(\omega)$ is not directly accessible by optical spectroscopy experiment. However, since $\sigma_1^{res}(\omega)$ and $\sigma_2^{res}(\omega)$ form a Kramers-Kronig pair one can get $\sigma_2^{res}(\omega)$ from the measured $\sigma_1^{res}(\omega)$ using a Kramers-Kronig relation\cite{wooten}. Once $\sigma_2^{res}(\omega)$ is obtained the superfluid plasma frequency can be estimated using the following equation,
\begin{equation}\label{eq9}
\omega_{sp}^2 = 4 \pi \omega [\sigma_2^{SC}(\omega) - \sigma_2^{res}(\omega)].
\end{equation}
One can also rewrite the equation above using Eq. (8) as
\begin{equation}\label{eq10}
\omega_{sp}^2 = \omega^2[\epsilon_H - \epsilon_1^{SC}(\omega)] - 4\pi \omega \sigma_2^{res}(\omega).
\end{equation}
The equation above can be reduced to a more practical equation, $\omega_{sp}^2 = \lim_{\omega \rightarrow 0}[-\omega^2 \epsilon_1^{SC}(\omega)]$ since $\epsilon_H$ is negligibly small compared with $\epsilon_1^{SC}(0)$ and $\sigma_2^{res}(\omega)$ is a regular function at $\omega =$ 0. One also can get independently the superfluid plasma frequency ($\omega_{sp}$) by taking advantage of the missing spectral weight, which is widely used and known as the Ferrell-Glover-Tinkham (FGT) sum rule\cite{glover:1956,ferrell:1958,hwang:2007a}.

\section{Results of model calculation and discussions}

\begin{figure}[!htbp]
  \vspace*{-0.3 cm}%
  \centerline{\includegraphics[width=4.0 in]{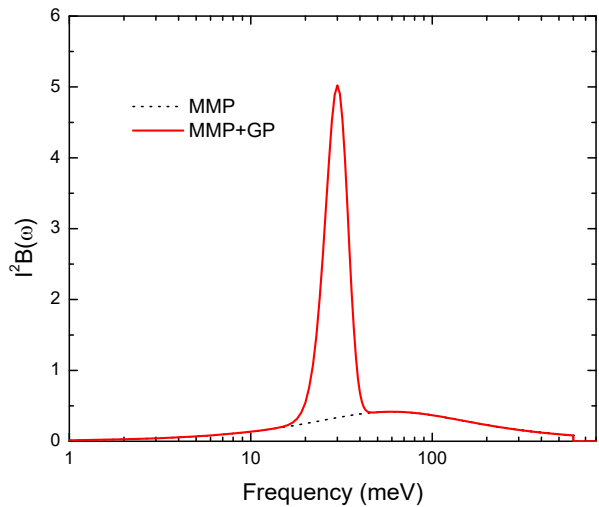}}%
  \vspace*{-0.3 cm}%
\caption{(Color online) Two typical input model electron-boson spectral density functions, $I^2B(\omega)$: the MMP and MMP+GP models (refer to the text).}
 \label{fig1}
\end{figure}

For our theoretical calculations we focus on four (normal, half pseudogap (GP), full pseudogap (GP), and $d$-wave superconducting (SC)) phases with two typical input $I^2B(\omega)$ models\cite{hwang:2006,carbotte:2011}. One typical $I^2B(\omega)$ is modelled for antiferromagnetic spin-fluctuations at high temperature, which consists of the broad Millis, Monien, and Pines (MMP) mode\cite{millis:1990} alone and the other $I^2B(\omega)$ is modelled for antiferromagnetic spin-fluctuations at low temperature, which consists of two components: one sharp Gaussian peak (GP) and the broad MMP mode\cite{millis:1990}. It is worth to be noted that the temperature-dependent evolution of $I^2B(\omega)$ is well-established by various spectroscopy techniques including optical spectroscopy\cite{hwang:2004,hwang:2006,hwang:2007,dai:1999,fong:2000,carbotte:2011,johnson:2001,valla:2007,zhang:2008}. In Fig. \ref{fig1} we display the two typical (MMP and MMP+GP) model $I^2B(\omega)$. The MMP+GP model is in the red solid line in the figure and consists of two terms which can be written as $I^2B(\omega) = A_s \: \omega/(\omega^2+\Omega_{sf}^2) +  A_p/[\sqrt{2\pi(D/2.35)^2}]\exp\{-(\omega-\Omega_0)^2/[2(D/2.35)^2]\}$ where $A_s$ (= 50 meV) and $\Omega_{sf}$ (= 60 meV) are, respectively, the amplitude and the characteristic frequency of the MMP mode and $A_p$ (= 50 meV), $D$ (= 10 meV), and $\Omega_0$ (= 30 meV) are, respectively, the amplitude, width, and center frequency of the sharp GP mode. The MMP model is in the black dotted line in the figure and consists of a single term as $I^2B(\omega) = A_s \: \omega/(\omega^2+\Omega_{sf}^2)$ with the same $A_s$ and $\Omega_{sf}$ values. The area under the MMP mode is known as $A_s/2\: \ln|(\omega_{sf}^2+\omega_c^2)/\omega_{sf}^2|$\cite{hwang:2011}, where $\omega_c$ is a cutoff frequency, in our case 600 meV. Therefore, the areas under the MMP and MMP+GP curves are 115.4 and 165.4 meV, respectively. The correlation constant ($\lambda$) of $I^2B(\Omega)$ can be defined by $\lambda \equiv 2\int_0^{\omega_c} d\Omega\: I^2B(\Omega)/\Omega$. The calculated correlation constants are 2.42 and 5.83 for the MMP and MMP+GP models, respectively.

\begin{figure}[!htbp]
  \vspace*{-0.3 cm}%
  \centerline{\includegraphics[width=3.5 in]{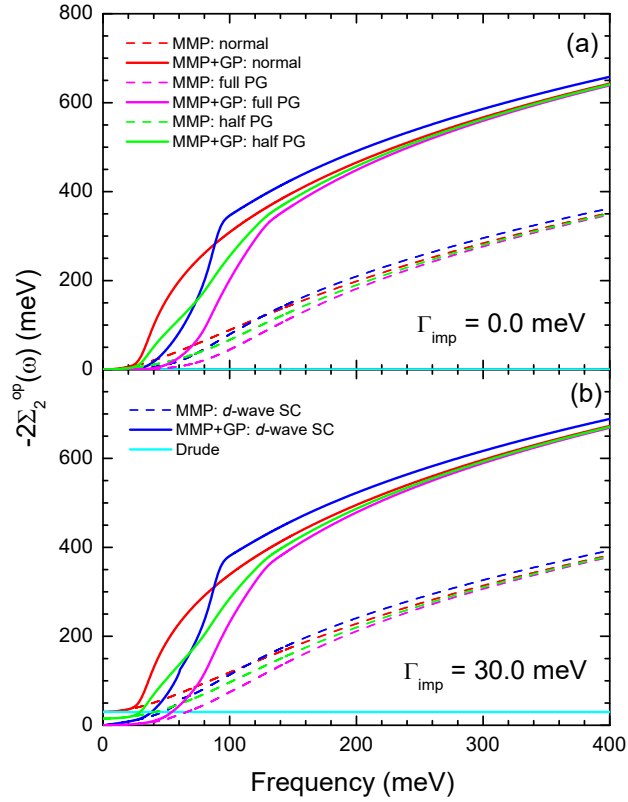}}%
  \vspace*{-0.3 cm}%
\caption{(Color online) The calculated imaginary parts of the optical self-energy for the four (normal, half PG, full PG, and $d$-wave SC) phases and two different impurity scattering rate cases ($\Gamma_{imp} =$ 0.0 and 30.0 meV).}
 \label{fig2}
\end{figure}

In Fig. \ref{fig2} we display calculated imaginary parts of the optical self-energy for the four (normal, half PG, full PG, and $d$-wave SC) phases with two different input $I^2B(\omega)$ (MMP and MMP+GP models) and two different impurity scattering rates, $\Gamma_{imp} =$ 0.0 meV (upper frame a) and 30.0 meV (lower frame b). Here we note that for the PG phase we have two different pseudogap depths, $N_0 =$ 0.5 (half PG) and 0.0 (full PG). For the PG and $d$-wave SC phases, as we can expect from Eq. (5) the significant impurity effects appear in low frequency region below twice of the maximum superconducting gap ($2\Delta_0 =$ 60 meV) and twice of the pseudogap ($2\Delta_{PG} =$ 100 meV). Above these energy scales each calculated imaginary optical self-energy for $\Gamma_{imp} =$ 30 meV roughly is increased by 30 meV compared with the corresponding self-energy with $\Gamma_{imp} =$ 0.0 meV. For the normal phase since the impurity scattering is a (frequency-independent) constant, 30 meV, the imaginary self-energy is vertically shifted by 30 meV in the whole spectral range compared with the corresponding self-energy with $\Gamma_{imp} =$ 0.0 meV. While the PG imaginary optical self-energy is always lower than the normal one the $d$-wave SC self-energy is lower in low frequency region and becomes higher in high frequency region than the normal one. The location of the crossing point between the normal and SC imaginary self-energies seems to depend on the shape of the model $I^2B(\omega)$. We note that the half PG optical self-energy is in between the normal and full PG self-energies. At high frequency region all imaginary self-energies for the same model $I^2B(\omega)$ are getting merged together. It is worth to be noted that, in principle, the imaginary self-energy is eventually saturated to $2\pi \times$the area under $I^2B(\omega)$\cite{hwang:2008a}. The saturated values for the MMP and MMP+GP models are 724 meV and 1038 meV, respectively, for $\Gamma_{imp} =$ 0.0 meV. Comparing the imaginary optical self-energy of the MMP+GP model with that of the MMP model we can see that the sharp GP gives a strong jump in the imaginary self-energy, which appears as a step-like feature\cite{hwang:2004}. From this step one can easily estimate the strength of the sharp GP mode. Interestingly, the shape and onset frequency of the step seem to depend on the (normal, half PG, full PG, and $d$-wave SC) phases of materials. The characteristic energy scales appear near the step-like feature (refer to Fig. \ref{fig4}). The imaginary part of the optical self-energy is also known as the optical scattering rate, $1/\tau^{op}(\omega) = -2\Sigma^{op}_2(\omega)$, which is the scattering rate between interacting electrons by exchanging the mediating bosons at a given frequency, $\omega$.

\begin{figure}[!htbp]
  \vspace*{-0.3 cm}%
  \centerline{\includegraphics[width=3.5 in]{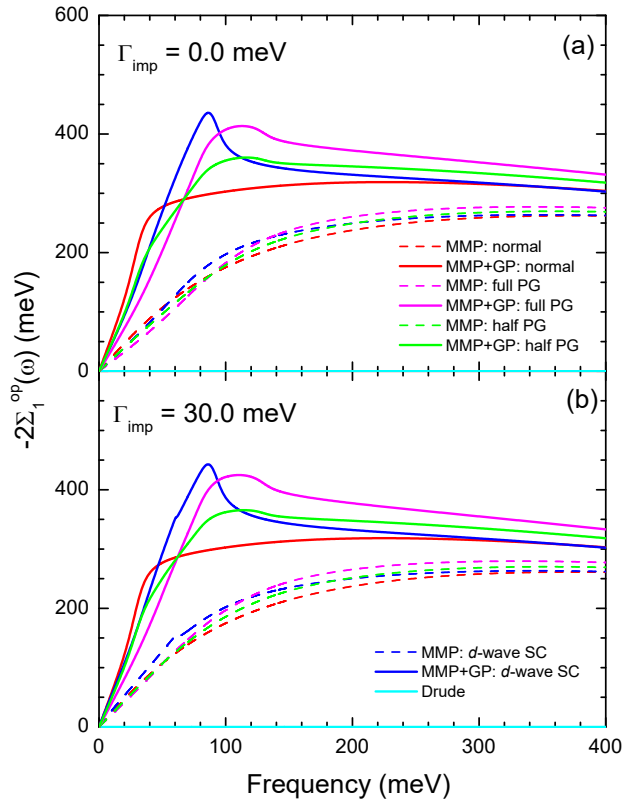}}%
  \vspace*{-0.7 cm}%
\caption{(Color online) The corresponding real parts of the optical self-energy for the  four (normal, half PG, full PG, and $d$-wave SC) phases and two different impurity scattering cases ($\Gamma_{imp} =$ 0.0 and 30.0 meV).}
 \label{fig3}
\end{figure}

In Fig \ref{fig3} we display corresponding real parts of the optical self-energy to the imaginary parts displayed in Fig. \ref{fig2}. We note that these real parts are obtained from the imaginary parts in a wide spectral range from 0 to 7000 meV using the Kramers-Kronig relation, Eq. (6). Therefore, in principle, the real part should carry additional independent information. The real self-energy shows the resonance (potential) energy between interacting electrons by exchanging the mediating bosons at a given frequency, $\omega$. The (energy) width of the resonance potential is determined by the imaginary self-energy. This real self-energy can carry the information of the effective mass of charge carriers caused by the interaction, i.e. $-2\Sigma^{op}_1(\omega) = [m_{op}^*(\omega)/m_b -1]\omega$. Interestingly, there are isosbestic points among the normal, half PG, and full PG phases for both MMP and MMP+GP models and both $\Gamma_{imp} =$ 0.0 and 30.0 meV even though the positions of all the isosbestic points are located at different frequencies. It is worth to be noted that for the normal phase, the real part of the optical self-energy is independent of the impurity scattering rate since the scattering rate is a constant; the Kramers-Kronig transformation of a constant is always zero. For the broad MMP model case, all self-energies are quite similar to one another, which means that the different material phases (normal, half PG, full PG, and SC phases) are not very well-resolved. In contrast, for the MMP+GP model case, the sharp Gaussian peak allows us to clearly resolve the material phases (normal, half PG, full PG, and SC phases), particularly in low frequency region where the peak position ($\Omega_0$), the PG ($2\Delta_{PG}$), and the maximum SC gap ($2\Delta_0$) exist. Therefore, if the Gaussian peak shows some doping- and temperature-dependent evolutions one can easily observe the doping- and temperature-dependent evolutions of the intensity and/or the width of the Gaussian peak as reported by previous studies\cite{hwang:2004,hwang:2006,hwang:2007}. The Gaussian peak gives a sharp peak in the $d$-wave SC phase, a well-defined round peak in the two PG phases\cite{hwang:2008}, and a logarithmic singularity in the normal phase\cite{carbotte:2005}. In Fig. \ref{fig4} we display derivatives of the imaginary part (upper frame) and real part (lower frame) of the optical self-energy for the MMP+GP model and $\Gamma_{imp} =$ 30.0 meV. In the figure, we can see that various characteristic energy scales ($\Omega_0$, $\Delta_{PG}$, and $2\Delta_0$) and their combinations appear as peaks or dips. If the Gaussian peak becomes sharper the features will become better-defined and give more accurate values.

\begin{figure}[!htbp]
  \vspace*{-0.3 cm}%
  \centerline{\includegraphics[width=3.5 in]{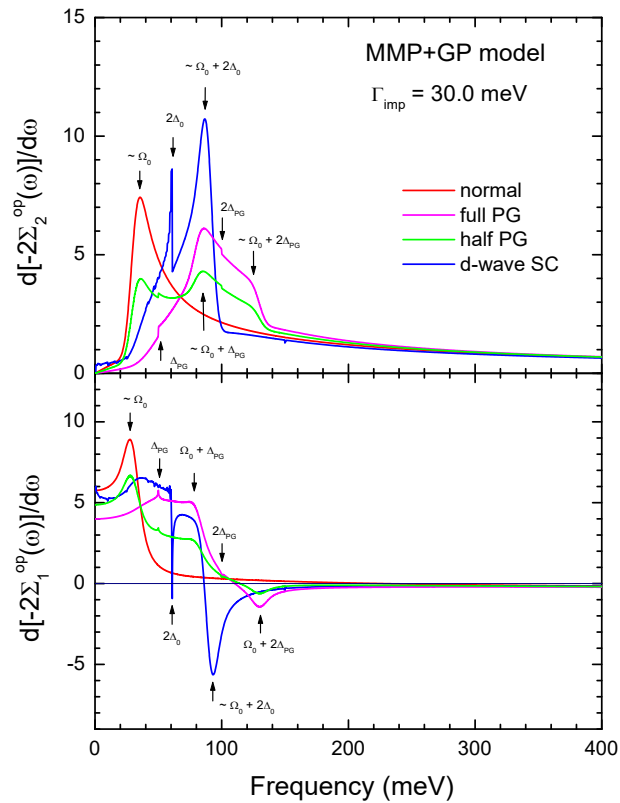}}%
  \vspace*{-0.7 cm}%
\caption{(Color online) Derivatives of the imaginary (upper frame) and real (lower frame) parts of the calculated optical self-energy for the MMP+GP model with $\Gamma_{imp} =$ 30.0 meV.}
 \label{fig4}
\end{figure}

Now we consider the complex optical conductivities, the complex dielectric functions, and reflectance spectra which can be obtained from the calculated complex optical self-energies. Using equations which we previously introduced in the "Theoretical Formalism" section, we obtained the complex optical conductivity (Eq. (7)) and the complex dielectric function (Eq. (8)), the loss function (Eq. (9)) and the normal incidence reflectance (Eq. 10).

\begin{figure}[!htbp]
  \vspace*{-0.3 cm}%
  \centerline{\includegraphics[width=4.0 in]{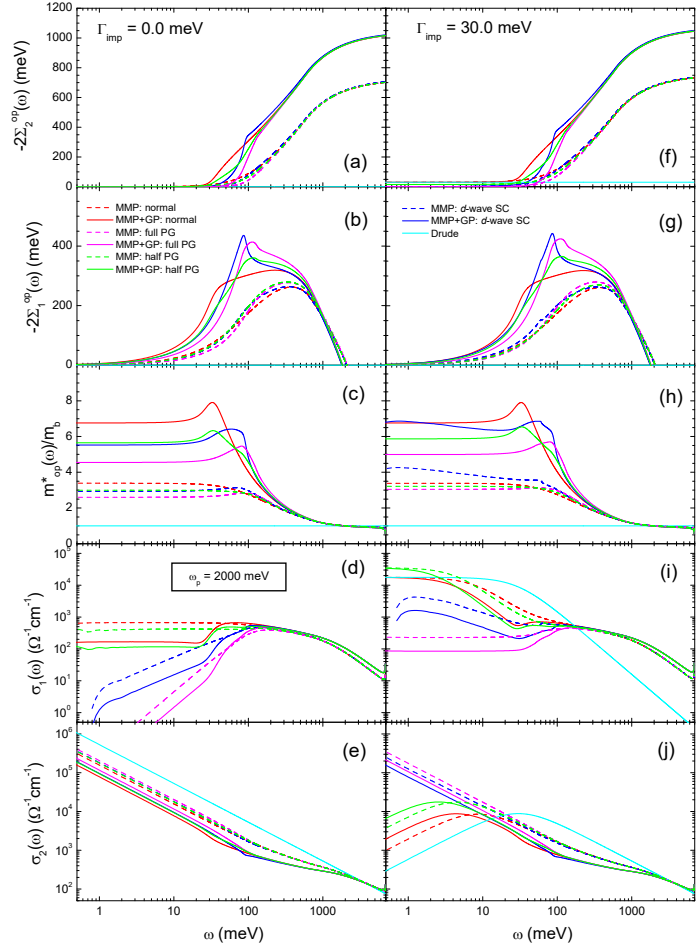}}%
  \vspace*{-0.5 cm}%
\caption{(Color online) Calculated complex optical self-energy, mass enhancement factor, and complex optical conductivity for normal, half PG, full PG, and $d$-wave SC states with two different impurity scattering rates, $\Gamma_{imp} =$ 0.0 (left column) and 30.0 meV (right column).}
 \label{fig5}
\end{figure}

In Fig. \ref{fig5} we display the obtained complex optical conductivities along with the corresponding complex optical self-energies in a wide spectral range from 0 to 7000 meV. The left-hand and right-hand side columns are for $\Gamma_{imp} =$ 0.0 meV and $\Gamma_{imp} =$ 30.0 meV cases, respectively. In an order from the top to the bottom we display the imaginary and real parts of the optical self-energy ($-2\Sigma^{op}_2(\omega)$ and $-2\Sigma^{op}_1(\omega)$), the mass enhancement function ($m^*_{op}(\omega)/m_b$), and the real and imaginary parts of the optical conductivity ($\sigma_1(\omega)$ and $\sigma_2(\omega)$). For getting the complex optical conductivity using the extended Drude mode (or Eq. (7)) we used 2000 meV for the plasma frequency, $\omega_p$. For comparisons we provided the corresponding optical quantities for a free electron system (or the Drude mode) with the same plasma frequency of 2000 meV and the impurity scattering rates in the cyan solid lines.

First, we describe the quantities on the left-hand side column of Fig. \ref{fig5} (i.e., for $\Gamma_{imp}=$ 0.0 meV case) from the top to the bottom. In Fig. \ref{fig5}(a) and \ref{fig5}(b) the values of the imaginary optical self-energy at 7000 meV are 702 meV and 1016 meV for the MMP and MMP+GP models, respectively, which are close to the estimated saturation values (i.e., $2\pi \times$the area under the $I^2B(\omega)$ curve) 725 and 1039 meV, respectively. As we expected, both real and imaginary parts of the optical self-energy for the Drude mode are completely zero due to no correlations and zero impurity scattering rate between electrons. In Fig. \ref{fig5}(c) the mass enhancement function at zero frequency shows a strong phase-dependence; the value for the normal phase is the largest and the value for the full PG phase is the smallest. The mass enhancement function for the Drude mode is exactly 1.0 in the whole frequency range. In Fig. \ref{fig5}(d) the real part of the optical conductivity of the Drude mode does not appear in a finite spectral range; since $\Gamma_{imp} =$ 0.0 meV, the whole spectral weight ($\omega_p^2/8$) is concentrated at zero frequency and appears as a Dirac delta function, $(\omega_p^2/8)\delta(\omega)$. Therefore, no spectral weight appears in a finite frequency range. The spectral weight of the delta function is called a missing spectral weight\cite{glover:1956,ferrell:1958}. However, this missing spectral weight can be seen in the imaginary part of the optical conductivity as $1/\omega$ frequency-dependence at low frequency region near zero frequency as we can see in Fig. \ref{fig5}(e). In fact, the parallel lines with the same slope of -1 in a log-log plot in Fig. \ref{fig5}(e) show the missing spectral weights for all cases. It should be noted that, in this formalism, a normal phase with $\Gamma_{imp} =$ 0.0 meV at $T =$ 0 is identical to a zero-gap superconducting phase.

\begin{figure}[!htbp]
  \vspace*{-0.3 cm}%
  \centerline{\includegraphics[width=4.0 in]{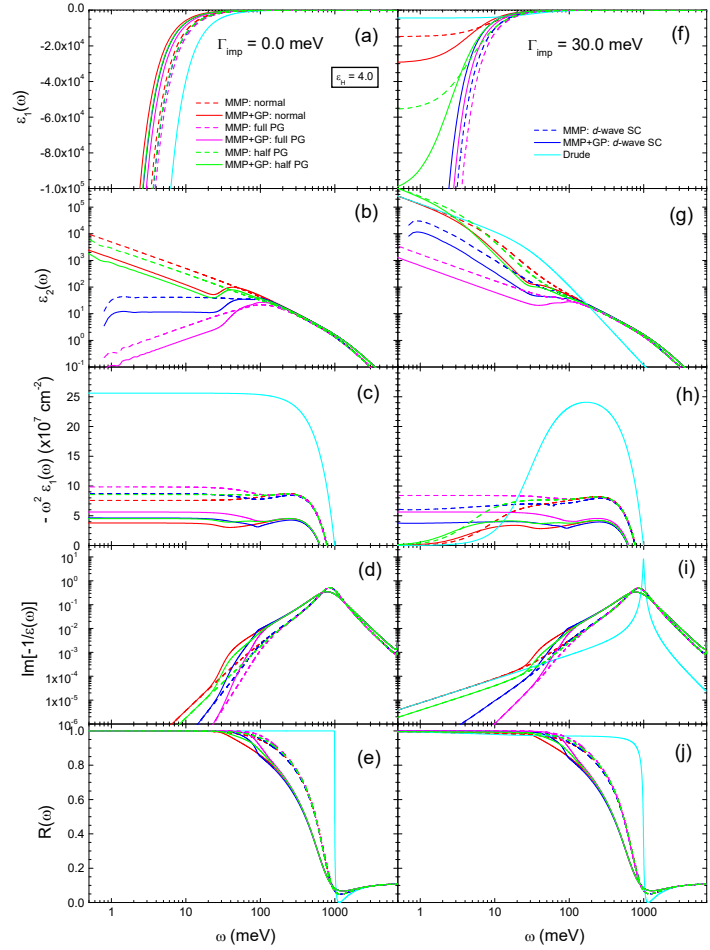}}%
  \vspace*{-0.5 cm}%
\caption{(Color online) Calculated complex dielectric functions, $-\omega^2\epsilon_1(\omega)$ quantities, the loss functions, and reflectance spectra for normal, half PG, full PG, and $d$-wave SC states with two different impurity scattering rates, $\Gamma_{imp} =$ 0.0 (left column) and 30.0 meV (right column).}
 \label{fig6}
\end{figure}

Now we describe the corresponding quantities on the right-hand side column of Fig. \ref{fig5} (i.e., for $\Gamma_{imp}=$ 30.0 meV case) from the top to the bottom. In Fig. \ref{fig5}(f) and \ref{fig5}(g) the values of the imaginary optical self-energy at 7000 meV are now 732 meV and 1046 meV for the MMP and MMP+GP models, respectively; compared with the corresponding values for the $\Gamma_{imp}=$ 0.0 meV case, both increase by the impurity scattering rate, 30 meV. Even though the imaginary self-energy for the Drude mode is a constant, 30 meV, the real part is still zero since a Kramers-Kronig transformation of a constant is always zero. In Fig. \ref{fig5}(h) the mass enhancement functions of the $d$-wave SC phase for both MMP and MMP+GP models are enhanced significantly at low frequency region by the increased impurity scattering while those of the normal and PG phases remain more or less the same values. The mass enhancement for the Drude mode is still exactly 1.0. In Fig. \ref{fig5}(i) and \ref{fig5}(j), since $\Gamma_{imp} =$ 30.0 meV, $\sigma_1(\omega)$ of the Drude mode appears in the wide spectral range as a Lorentzian curve which is located at zero frequency with a width (30 meV). For the normal and half PG phases, no missing spectral weights exist any more. However, for the full PG and $d$-wave SC phases some missing spectral weights still exist. We can clearly see them in $\sigma_2(\omega)$ at low frequency region; while for the Drude mode the normal and half PG phases $\sigma_2(0) =$ 0, for the full PG and $d$-wave SC phases $\sigma_2(0) \neq$ 0. Therefore, in the full PG and $d$-wave SC cases we still observe the remaining missing spectral weights in $\sigma_2(\omega)$ at low frequency region, which appear as parallel lines with the same slope of -1 in the log-log plot. In the real and imaginary parts of the optical conductivity for the MMP+GP model we also can clearly see the various characteristic energy scales (refer to Fig. \ref{fig4}).

In Fig. \ref{fig6} we display the obtained complex dielectric functions and related quantities including reflectance spectra in a wide spectral range from 0 to 7000 meV. The left-hand and right-hand side columns are for $\Gamma_{imp} =$ 0.0 meV and $\Gamma_{imp} =$ 30.0 meV cases, respectively. In an order from the top to the bottom we display the imaginary and real parts of the dielectric function ($\epsilon_1(\omega)$ and $\epsilon_2(\omega)$), a quantity associated with the superfluid density or the superfluid plasma frequency ($-\omega^2\epsilon_1(\omega)$), the energy loss function ($Im[-1/\tilde{\epsilon}(\omega)$), and the reflectance ($R(\omega)$). For getting the complex dielectric functions using Eq. (8) we used 4.0 for the background dielectric constant, $\epsilon_H$. For comparisons we also provided the corresponding optical quantities for a free electron system with the same plasma frequency of 2000 meV and the same background dielectric constant of 4.0 in the cyan solid lines.

Here we describe the quantities on the left-hand side column of Fig. \ref{fig6} (i.e., for $\Gamma_{imp}=$ 0.0 meV case) from the top to the bottom. In Fig. \ref{fig6}(a) and \ref{fig6}(b) we display the real and imaginary parts of the dielectric function. These quantities basically carry the same information as the imaginary and real parts of the optical conductivity, respectively, i.e., $\epsilon_1(\omega) = \epsilon_H-[4\pi\sigma_2(\omega)]/\omega$ and $\epsilon_2(\omega) = 4 \pi \sigma_1(\omega)/\omega$. All $\epsilon_1(\omega)$ data show $-1/\omega^2$ dependence in low frequency region, which is directly related to the $1/\omega$ dependence in the corresponding $\sigma_2(\omega)$ (refer to Fig. \ref{fig5}(e)). In Fig. \ref{fig6}(c) the quantity $-\omega^2\epsilon_1(\omega)$ is associated with the superfluid plasma frequency ($\omega_{sp}$), i.e. $\omega_{sp}^2 = \lim_{\omega\rightarrow 0}[-\omega^2\epsilon_1(\omega)]$. Since $\Gamma_{imp} =$ 0.0, all phases including the Drude model have some amounts of the superfluid. Here we can see crossing points of $\epsilon_1(\omega)$ with the horizontal axis, which are associated with the effective plasma frequencies. In Fig. \ref{fig6}(d) the energy loss functions, which can be directly measured by electron energy loss spectroscopy (EELS), are displayed. The obtained energy loss functions show peaks near the effective plasma frequencies and, additionally, some fine features in low frequency region below 200 meV which are associated with the characteristics of $I^2B(\omega)$, the PG, and the maximum SC gap. In Fig. \ref{fig6}(e), we display corresponding reflectance spectra. It is worth to be noted that reflectance spectra can be directly measured using optical spectroscopy technique. In high frequency region above the plasma frequency the reflectance saturates to $|(\sqrt{\epsilon_H}-1)/(\sqrt{\epsilon_H}+1)|^2 \cong$ 0.11.

We also describe the quantities on the right-hand side column of Fig. \ref{fig6} (i.e., for $\Gamma_{imp}=$ 30.0 meV case) from the top to the bottom. In Fig. \ref{fig6}(f) $\epsilon_1(\omega)$ data for the normal, half PG, and the Drude phases are deviated from the $-1/\omega^2$ dependent behavior. In Fig. \ref{fig6}(g), in general, $\epsilon_2(\omega)$ looks very similar to $\sigma_1(\omega)$ since they are connected with a very simple relation, $\epsilon_2(\omega) = 4 \pi \sigma_1(\omega)/\omega$. In Fig. \ref{fig6}(h) only full PG and $d$-wave SC phases show the superfluid density. We clearly observe the characteristic energy scales associated with $I^2B(\omega)$, the pseudogap, and the maximum SC gap in the spectra. In Fig. \ref{fig6}(i) now we clearly see the Drude mode, which appears as a sharp peak at the effective plasma frequency, $\omega_p/\sqrt{\epsilon_{H}} =$ 1000 meV. In Fig. \ref{fig6}(j) we can clearly observe the related characteristic features in the calculated reflectance spectra. Another important information that one may be able to get from these calculated reflectance spectra is the model-dependent extrapolation to zero frequency for the Kramers-Kronig analysis of measured reflectance spectrum. There are some useful extrapolations for this purpose\cite{quijada:1999}, as an example, the Hagen-Rubens relation (i.e., $1-R(\omega) = A\sqrt{\omega}$) for simple metals, where $A$ is an adjustable parameter. In low frequency region $1-R(\omega)$ looks very similar to $-2\Sigma^{op}_2(\omega)$ (or $1/\tau^{op}(\omega)$)\cite{hwang:2015a} since, in general, in the relaxation region (i.e. $1/\tau_{imp} < \omega \ll \omega_p$) $1-R(\omega) \cong [2/\omega_p][1/\tau^{op}(\omega)]$.

\section{Comparison of the calculated data with measured data}

\begin{figure}[!htbp]
  \vspace*{-0.0 cm}%
  \centerline{\includegraphics[width=4.0 in]{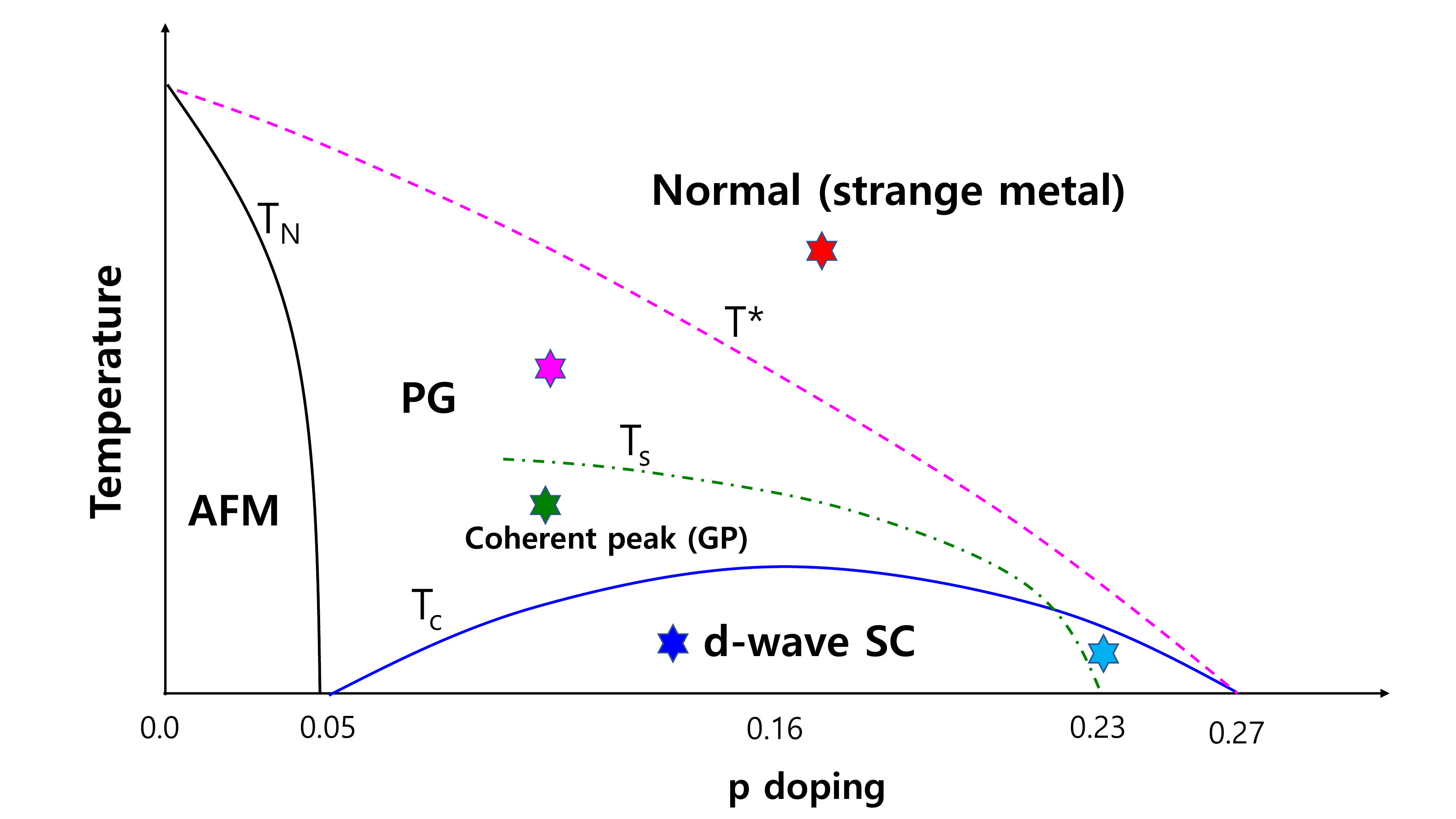}}%
  \vspace*{-0.0 cm}%
\caption{(Color online) Generic $T$-$p$ phase diagram of hole-doped cuprates: five designated regions are marked with stars of David.}
 \label{fig7}
\end{figure}

Now we compare the calculated optical quantities with measured optical data of Bi$_2$Sr$_2$CaCu$_2$O$_{8+\delta}$ (Bi-2212) at various doping levels and temperatures by considering the generic phase diagram of hole-doped cuprates shown in Fig. \ref{fig7}. There are four characteristic temperatures, $T_N$, $T^*$, $T_s$ and $T_c$, which are the Neel temperature, the pseudogap temperature, the coherent temperature, and the superconducting transition temperature, respectively. $T_s$ is also known as the onset temperature of the sharp GP mode in the $I^2B(\omega)$\cite{basov:2005}, which seems to be closely related to the onset temperature of the magnetic resonance mode\cite{dai:1999}. We focus on five different designated regions in the phase diagram marked with stars of David. We assigned those five regions as follows: the region marked with the red star is the normal (or strange metal) phase above $T^*$, which can be modelled with the MMP model in the normal phase\cite{carbotte:2011}. The region marked with the magenta star is the PG phase located between $T^*$ and $T_s$, which can be modelled with the MMP model in the half PG phase. The region marked with the olive star is also the PG region but located between $T_s$ and $T_c$, which can be modelled with the MMP+GP model in the full PG phase. The region marked with the blue star is the $d$-wave SC dome located below both $T_c$ and $T_s$, which can be modelled with the MMP+GP model in the $d$-wave SC phase. The region marked with the cyan star is also the $d$-wave SC dome but located below $T_c$ and above $T_s$, which can be modelled with the MMP model in the $d$-wave SC phase. These assignments are based on experimental results obtained by various spectroscopy techniques.\cite{hwang:2004,hwang:2006,hwang:2007,dai:1999,fong:2000,zasadzinski:2006,zasadzinski:2011,carbotte:2011,johnson:2001}

\begin{figure}[!htbp]
  \vspace*{-0.3 cm}%
  \centerline{\includegraphics[width=4.0 in]{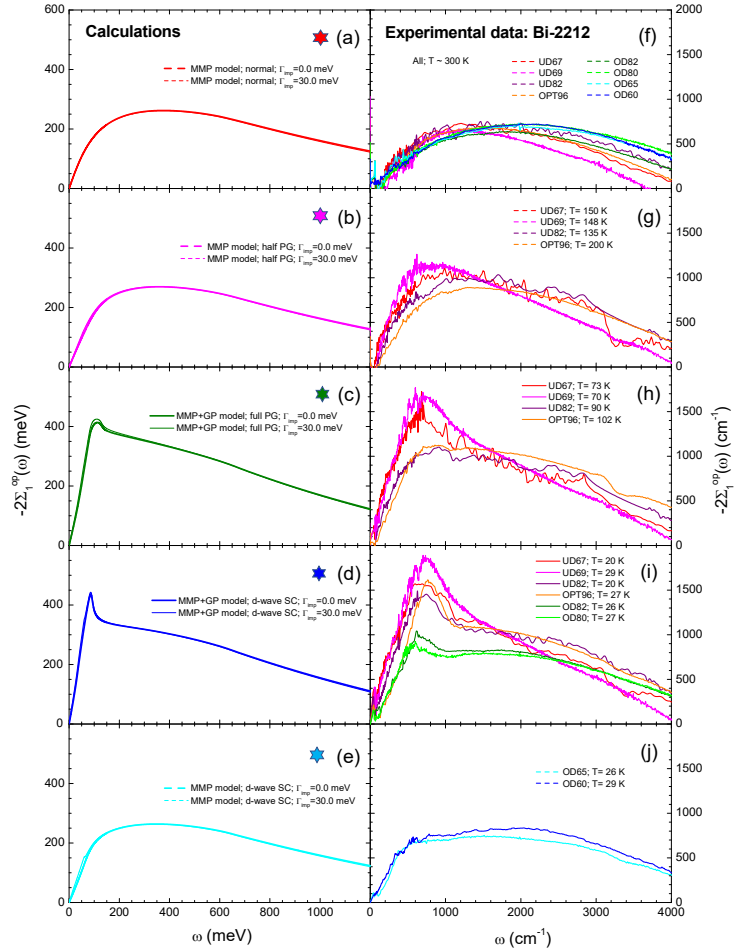}}%
  \vspace*{-0.3 cm}%
\caption{(Color online) Comparison of the calculated real self-energies (left column) of the five marked regions in the phase diagram with measured real self-energies of Bi-2212 (right column) reproduced from a published paper\cite{hwang:2007a} in the marked regions.}
 \label{fig8}
\end{figure}

For comparing the calculated data with experimentally measured ones, we choose the real parts of the optical self-energy among all quantities considered so far. We think that this quantity is most reliable for the comparison since it is not very sensitive to a temperature-induced impurity scattering rate. In the left-hand side column of Fig. \ref{fig8} we display the calculated real parts of the optical self-energy in the five designated regions of the phase diagram from the top to the bottom (a-e). In the right-hand side column of Fig. \ref{fig8} we display measured real optical self-energies of eight Bi-2212 samples in different doping levels at various temperatures in the five marked regions of the phase diagram from the top to the bottom (f-j). We note that the measured real self-energies of Bi-2212 have been published in a literature\cite{hwang:2007a}. The doping levels of the eight Bi-2212 samples are extended from underdoped to overdoped. We denote the eight samples in an order from underdoped to overdoped as UD67, UD69, UD82, OPT96, OD82, OD80, OD65, and OD60, where UD, OPT, and OD stand for underdoped, the optimally doped, and overdoped, respectively, and the number indicates the superconducting transition temperature ($T_c$) of each sample. It is worth to be noted that the calculated data do not include thermal effects. It also should be noted that we used different energy units and scales in the plots for the calculated and measured data sets. We describe each row of Fig. \ref{fig8} more in detail as follows:

Row 1: In Fig. \ref{fig8}(a) we display the calculated real self-energies in the region marked with the red star (i.e., the MMP model in the normal phase) for both $\Gamma_{imp} =$ 0.0 meV and 30.0 meV cases. In Fig. \ref{fig8}(f) we display the measured real self-energies of all eight Bi-2212 samples at $\sim$300 K, i.e., in the region marked with the red star (or the normal region). The MMP modes of the measured Bi-2212 samples seem to show doping-dependencies; as the doping increases the characteristic frequency ($\omega_{sf}$) of the MMP mode increases and the MMP mode spreads more widely to high frequency. We note that even though according to the phase diagram in a literature\cite{basov:2005} the pseudogap temperatures ($T^*$) of all three underdoped Bi-2212 samples are larger than 300 K we display the self-energy spectra at $\sim$ 300 K in the normal region marked with the red star since $T^*$ temperatures in the underdoped region of the phase diagram \cite{basov:2005} show large error bars.

Row 2: In Fig. \ref{fig8}(b) we display the calculated real self-energies in the region marked with the magenta star (i.e., the MMP model in the half PG phase) for both $\Gamma_{imp} =$ 0.0 meV and 30.0 meV cases. In Fig. \ref{fig8}(g) we display the measured real self-energies of UD67 at 150 K, UD69 at 148 K, UD82 at 135 K, and OPT96 at 200 K, which are located in the region marked with the magenta star according to the phase diagram in the literature\cite{basov:2005}.

Row 3: In Fig. \ref{fig8}(c) we display the calculated real self-energies in the region marked with the olive star (i.e. the MMP+GP model in the full PG phase) for both $\Gamma_{imp} =$ 0.0 meV and 30.0 meV cases. In Fig. \ref{fig8}(h) we display the measured real self-energies of UD67 at 73 K, UD69 at 70 K, UD82 at 90 K, and OPT96 at 102 K, which are located in the region marked with the olive star. It is worth to be noted that UD67 and UD69 show much stronger peaks than UD82 and OPT96; this observation might be understood by considering the doping-dependent size\cite{hufner:2008} of the PG and the temperature- and doping-dependent intensity of the GP mode\cite{hwang:2007,hwang:2011,hwang:2016a} as well.

Row 4: In Fig. \ref{fig8}(d) we display the calculated real self-energies in the region marked with the blue star (i.e. the MMP+GP model in the $d$-wave SC phase) for both $\Gamma_{imp} =$ 0.0 meV and 30.0 meV cases. In Fig. \ref{fig8}(i) we display the measured real self-energies of UD67 at 20 K, UD69 at 29 K, UD82 at 20 K, OPT96 at 27 K, OD82 at 26 K, and OD80 at 27 K, which are located in the region marked with the blue star. We clearly observe doping-dependent trend of $I^2B(\omega)$, which has been studied previously\cite{hwang:2004,hwang:2007}; as the doping increases the intensity of the GP mode decreases rapidly in the overdoped region of the phase diagram and the GP mode is red-shifted\cite{hwang:2004,hwang:2007,yang:2009}.

Row 5: In Fig. \ref{fig8}(e) we display the calculated real self-energies in the region marked with the cyan star (i.e. the MMP model in the $d$-wave SC phase) for both $\Gamma_{imp} =$ 0.0 meV and 30.0 meV cases. In Fig. \ref{fig8}(j) we display the measured real self-energies of OD65 at 26 K and OD60 at 29 K, which are located close to the region marked with the cyan star. Since the two highly overdoped samples (OD65 and OD60) show a very much suppressed Gaussian peak (GP) the two samples in the $d$-wave SC phase can be approximately described by the MMP mode alone\cite{hwang:2004,hwang:2007}.

This comparison shows that the calculated and measured data in each designated region of the phase diagram qualitatively agree well each other; the marked regions seem to be well-characterized by the specified models. Therefore, the simple MMP and MMP+GP models with the normal, half PG, full PG, and $d$-wave SC phases seem to reasonably describe the well-known phase diagram of hole-doped cuprates.

\section{Conclusions}

We obtained the complex optical self-energy in four different material phases (normal, half PG, full PG, and $d$-wave SC) with the two typical MMP and MMP+GP model electron-boson spectral density functions, $I^2B(\omega)$, using the generalized Allen's formulas and the Kramers-Kronig relation between the real and imaginary parts of the optical self-energy. We also obtained the corresponding complex optical conductivity from the obtained complex optical self-energy using the extended Drude model formalism. Furthermore, we got the other optical constants including reflectance spectrum using the relationships between optical constants. From this study we observed significant differences among the calculated various optical quantities with the same $I^2B(\omega)$ but in different material phases. We assigned the calculated optical quantities into designated regions of the generic phase diagram of hole-doped cuprates and compared the calculated real optical self-energies with measured optical self-energies of Bi-2212\cite{hwang:2007a} located in the same designated region of the phase diagram. From the comparison we concluded that the assignment is quite reasonable, and this optical data analysis can be a convincing method to expose the pairing force between electrons in the copper oxide superconductors and other superconducting materials as well. We hope that our results attract a lot of attentions from researchers in the field of superconductivity.

\ack

This paper was supported by Sungkyun Research Fund, Sungkyunkwan University, 2016.

\section*{References}
\bibliographystyle{unsrt}
\bibliography{bib}

\begin{thebibliography}{10}

\bibitem{mori:1965}
Hazime Mori.
\newblock A continued-fraction representation of the time-correlation
  functions.
\newblock {\em Prog. Theor. Phys.}, 34:399, 1965.

\bibitem{gotze:1972}
W.~G\"{o}tze and P.~W\"{o}lfle.
\newblock Homogeneous dynamical conductivity of simple metals.
\newblock {\em Phys. Rev. B}, 6:1226, 1972.

\bibitem{hwang:2004}
J.~Hwang, T.~Timusk, and G.~D. Gu.
\newblock High-transition-temperature superconductivity in the absence of the
  magnetic-resonance mode.
\newblock {\em Nature (London)}, 427:714, 2004.

\bibitem{carbotte:2005}
J.~P. Carbotte, E.~Schachinger, and J.~Hwang.
\newblock Boson structures in the relation between optical conductivity and
  quasiparticle dynamics.
\newblock {\em Phys. Rev. B}, 71:054506, 2005.

\bibitem{hwang:2007b}
J.~Hwang, E.~J. Nicol, T.~Timusk, A.~Knigavko, and J.~P. Carbotte.
\newblock High energy scales in the optical self-energy of the cuprate
  superconductors.
\newblock {\em Phys. Rev. Lett.}, 98:207002, 2007.

\bibitem{tanner:1992}
D.~B. Tanner and T.~Timusk.
\newblock {\em in {\it Physical Properties of high Temperature Superconductors
  III}, edited by D. M. Ginsberg (World Scientific, Singapore, 1992)}.

\bibitem{kotlier:2004}
Gabriel Kotliar and Dieter Vollhardt.
\newblock Strongly correlated materials: Insights from dynamical mean-field
  theory.
\newblock {\em Phys. Today}, 57:53, 2004.

\bibitem{webb:1986}
B.~C. Webb, A.~J. Sievers, and T.~Mihalisin.
\newblock Observation of an energy- and temperature-dependent carrier mass for
  mixed-valence \mbox{CePd$_3$}.
\newblock {\em Phys. Rev. Lett.}, 57:1951, 1986.

\bibitem{sulewski:1988}
P.~E. Sulewski, A.~J. Sievers, M.~B. Maple, M.~S. Torikachvili, J.~L. Smith,
  and Z.~Fisk.
\newblock Far-infrared absorptivity of \mbox{UPt$_3$}.
\newblock {\em Phys. Rev. B}, 38:5338, 1988.

\bibitem{thomas:1988}
G.~A. Thomas, J.~Orenstein, D.~H. Rapkine, M.~Capizzi, A.~J. Millis, R.~N.
  Bhatt, L.~F. Schneemeyer, and J.~V. Waszczak.
\newblock \mbox{Ba$_2$YCu$_3$O$_{7-\delta}$} electrodynamics of crystals with
  high reflectivity.
\newblock {\em Phys. Rev. Lett.}, 61:1313, 1988.

\bibitem{awasthi:1993}
A.~M. Awasthi, L.~Degiorgi, G.~Gruner, Y.~Dalichaouch, and M.~B.Maple.
\newblock Complete optical spectrum of \mbox{CeAl$_3$}.
\newblock {\em Phys. Rev. B}, 48:10692, 1993.

\bibitem{puchkov:1996}
A.~V. Puchkov, D.~N. Basov, and T.~Timusk.
\newblock The pseudogap state in high-\mbox{T$_c$} superconductors: an infrared
  study.
\newblock {\em J. Phys.: Cond. Matter}, 8:10049, 1996.

\bibitem{dordevic:2001}
S.~V. Dordevic, D.~N. Basov, N.~R. Dilley, E.~D. Bauer, and M.~B. Maple.
\newblock Hybridization gap in heavy fermion compounds.
\newblock {\em Phys. Rev. Lett.}, 86:684, 2001.

\bibitem{tran:2002}
P.~Tran, S.~Donovan, and G.~Gruner.
\newblock Charge excitation spectrum in \mbox{UPt$_3$}.
\newblock {\em Phys. Rev. B}, 65:205102, 2002.

\bibitem{collins:1989}
R.~T. Collins, Z.~Schlesinger, F.~Holtzberg, P.~Chaudhari, and C.~Feild.
\newblock Reflectivity and conductivity of \mbox{YBa$_2$Cu$_3$O$_7$}.
\newblock {\em Phys. Rev. B}, 39:6571, 1989.

\bibitem{carbotte:1990}
J.~P. Carbotte.
\newblock Properties of boson-exchange superconductors.
\newblock {\em Rev. Mod. Phys.}, 62:1027, 1990.

\bibitem{schachinger:2000}
E.~Schachinger and J.~P. Carbotte.
\newblock Coupling to spin fluctuations from conductivity scattering rates.
\newblock {\em Phys. Rev. B}, 62:9054, 2000.

\bibitem{hwang:2006}
J.~Hwang, J.~Yang, T.~Timusk, S.~G. Sharapov, J.~P. Carbotte, D.~A. Bonn,
  R.~Liang, and W.~N. Hardy.
\newblock \mbox{a}-axis optical conductivity of detwinned ortho-\mbox{II}
  \mbox{YBa$_2$Cu$_3$O$_{6.50}$}.
\newblock {\em Phys. Rev. B}, 73:014508, 2006.

\bibitem{hwang:2007}
J.~Hwang, T.~Timusk, E.~Schachinger, and J.~P. Carbotte.
\newblock Evolution of the bosonic spectral density of the high-temperature
  superconductor \mbox{Bi$_2$Sr$_2$CaCu$_2$O$_{8+\delta}$}.
\newblock {\em Phys. Rev. B}, 75:144508, 2007.

\bibitem{hwang:2015a}
Jungseek Hwang.
\newblock Reverse process of usual optical analysis of boson-exchange
  superconductors: impurity effects on {\it s}- and {\it d}-wave
  superconductors.
\newblock {\em J. Phys.: Condens. Matter}, 27:085701, 2015.

\bibitem{hwang:2016}
J.~Hwang.
\newblock Electron-boson spectral density function of correlated multiband
  systems obtained from optical data: \mbox{Ba$_0.6$K$_0.4$Fe$_2$As$_2$} and
  lifeas.
\newblock {\em J. Phys.: Condens. Matter}, 28:125702, 2016.

\bibitem{basov:2005}
D.~N. Basov and T.~Timusk.
\newblock {\em Rev. Mod. Phys}, 77:721, 2005.

\bibitem{dai:1999}
Pengcheng Dai, H.~A. Mook, S.~M. Hayden, G.~Aeppli, T.~G. Perring, R.~D. Hunt,
  and F.~Doguan.
\newblock The magnetic excitation spectrum and thermodynamics of
  high-\mbox{$T_c$} superconductors.
\newblock {\em Science}, 284:1344, 1999.

\bibitem{fong:2000}
H.~F. Fong, P.~Bourges, Y.~Sidis, L.~P. Regnault, J.~Bossy, A.~Ivanov, D.~L.
  Milius, I.~A. Aksay, and B.~Keimer.
\newblock Spin susceptibility in underdoped \mbox{YBa$_2$Cu$_3$O$_{6+x}$}.
\newblock {\em Phys. Rev. B}, 61:14773, 2000.

\bibitem{zasadzinski:2006}
J.~F. Zasadzinski, L.~Ozyuzer, L.~Coffey, K.~E. Gray, D.~G. Hinks, and
  C.~Kendziora.
\newblock Persistence of strong electron coupling to a narrow boson spectrum in
  overdoped \mbox(bi$_2$sr$_2$cacu$_2$o$_{8+\delta}$.

\bibitem{zasadzinski:2011}
O.~Ahmadi, L.~Coffey, J.~F. Zasadzinski, N.~Miyakawa, and L.~Ozyuzer.
\newblock Eliashberg analysis of tunneling experiments: Support for the pairing
  glue hypothesis in cuprate superconductors.
\newblock {\em Phys. Rev. Lett.}, 106:167005, 2011.

\bibitem{carbotte:2011}
J.~P. Carbotte, T.~Timusk, and J.~Hwang.
\newblock Bosons in high-temperature superconductors: an experimental survey.
\newblock {\em Reports on Progress in Physics}, 74:066501, 2011.

\bibitem{allen:1971}
P.~B. Allen.
\newblock Electron-phonon effects in the infrared properties of metals.
\newblock {\em Phys. Rev. B}, 3:305, 1971.

\bibitem{sharapov:2005}
S.~G. Sharapov and J.~P. Carbotte.
\newblock Effects of energy dependence in the quasiparticle density of states
  on far-infrared absorption in the pseudogap state.
\newblock {\em Phys. Rev. B}, 72:134506, 2005.

\bibitem{schachinger:2006}
E.~Schachinger, D.~Neuber, and J.~P. Carbotte.
\newblock Inversion techniques for optical conductivity data.
\newblock {\em Phys. Rev. B}, 73:184507, 2006.

\bibitem{mitrovic:1985}
B.~Mitrovic and M.~A. Fiorucci.
\newblock Effects of energy dependence in the electronic density of states on
  the far-infrared absorption in superconductors.
\newblock {\em Phys. Rev. B}, 31:2694, 1985.

\bibitem{hwang:2008}
J.~Hwang, J.~P. Carbotte, and T.~Timusk.
\newblock Evidence for a pseudogap in underdoped
  \mbox{Bi$_2$Sr$_2$CaCu$_2$O$_{8+\delta}$} and \mbox{YBa$_2$Cu$_3$O$_6.50$}
  from in-plane optical conductivity measurements.
\newblock {\em Phys. Rev. Lett.}, 100:177005, 2008.

\bibitem{hwang:2008b}
J.~Hwang, J.~P. Carbotte, and T.~Timusk.
\newblock Fermi surface arcs and the infrared conductivity of underdoped
  \mbox{YBa$_2$Cu$_3$O$_{6.5}$}.
\newblock {\em Euro. Phys. Lett.}, 82:27002, 2008.

\bibitem{wooten}
Frederick Wooten.
\newblock {\em Optical Properties of Solids}.
\newblock Academic, New York, 1972.
\newblock (Note: Key material on page 176).

\bibitem{glover:1956}
R.~E. Glover and M.~Tinkham.
\newblock Transmission of superconducting films at millimeter-microwave and far
  infrared frequencies.
\newblock {\em Phys. Rev.}, 104:844, 1956.

\bibitem{ferrell:1958}
R.~A. Ferrell and R.~E. Glover.
\newblock Conductivity of superconducting films: A sum rule.
\newblock {\em Phys. Rev.}, 109:1398, 1958.

\bibitem{hwang:2007a}
J.~Hwang, T.~Timusk, and G.~D. Gu.
\newblock Doping dependent optical properties of
  \mbox{Bi$_2$Sr$_2$CaCu$_2$O$_{8+\delta}$}.
\newblock {\em J. Phys.: Condens. Matter}, 19:125208, 2007.

\bibitem{millis:1990}
A.~J. Millis, H.~Monien, and D.~Pines.
\newblock Phenomenological model of nuclear relaxation in the normal state of
  \mbox{YBa$_2$Cu$_3$O$_7$}.
\newblock {\em Phys. Rev. B}, 42:167, 1990.

\bibitem{johnson:2001}
P.~D. Johnson, T.~Valla, A.~V. Fedorov, Z.~Yusof, B.~O. Wells, Q.~Li, A.~R.
  Moodenbaugh, G.~D. Gu, N.~Koshizuka, C.~Kendziora, Sha Jian, and D.~G. Hinks.
\newblock {\em Phys. Rev. Lett.}, 87:177007, 2001.

\bibitem{valla:2007}
T.~Valla, T.~E. Kidd, W.-G. Yin, G.~D. Gu, P.~D. Johnson, Z.-H. Pan, and A.~V.
  Fedorov.
\newblock {\em Phys. Rev. Lett.}, 98:167003, 2007.

\bibitem{zhang:2008}
Wentao Zhang, Guodong Liu, Lin Zhao, Haiyun Liu, Jianqiao Meng, Xiaoli Dong,
  Wei Lu, J.~S. Wen, Z.~J. Xu, G.~D. Gu, T.~Sasagawa, Guiling Wang, Yong Zhu,
  Hongbo Zhang, Yong Zhou, Xiaoyang Wang, Zhongxian Zhao, Chuangtian Chen,
  Zuyan Xu, and X.~J. Zhou.
\newblock {\em Phys. Rev. Lett.}, 100:107002, 2008.

\bibitem{hwang:2011}
J.~Hwang.
\newblock Electron-boson spectral density function of underdoped
  \mbox{Bi$_2$Sr$_2$CaCu$_2$O$_{8+\delta}$} and \mbox{YBa$_2$Cu$_3$O$_{6.50}$}.
\newblock {\em Phys. Rev. B}, 83:014507, 2011.

\bibitem{hwang:2008a}
J.~Hwang, J.~Yang, J.~P. Carbotte, and T.~Timusk.
\newblock Manifestation of the pseudogap in ab-plane optical characteristics.
\newblock {\em J. Phys. Condens. Matter}, 20:295215, 2008.

\bibitem{quijada:1999}
M.~A. Quijada, D.~B. Tanner, R.~J. Kelley, M.~Onellion, H.~Berger, and
  G.~Margaritondo.
\newblock Anisotropy in the ab-plane optical properties of
  \mbox{Bi$_2$Sr$_2$CaCu$_2$O$_8$} single-domain crystals.
\newblock {\em Phys. Rev. B}, 60:14917, 1999.

\bibitem{hufner:2008}
S~H\"{u}fner, M~A Hossain, A~Damascelli, and G~A Sawatzky.
\newblock {\em Reports on Progress in Physics}, 71:062501, 2008.

\bibitem{hwang:2016a}
Jungseek Hwang.
\newblock Intrinsic temperature-dependent evolutions in the electron-boson
  spectral density obtained from optical data.
\newblock {\em Scientific Reports}, 6:23647, 2016.

\bibitem{yang:2009}
J.~Yang, J.~Hwang, E.~Schachinger, J.~P. Carbotte, R.~P. S.~M. Lobo, D.~Colson,
  A.~Forget, and T.~Timusk.
\newblock Exchange boson dynamics in cuprates: Optical conductivity of
  \mbox{HgBa$_{2}$CuO$_{4+\delta}$}.
\newblock {\em Phys. Rev. Lett.}, 102:027003, 2009.

\end{thebibliography}

\end{document}